# Effects of Shape on Interaction Dynamics of Tetrahedral Nanoplastics and the Cell Membrane

Xin Yong[1*], Ke Du[2]

[1] Department of Mechanical Engineering, Binghamton University, Binghamton, NY 13902, USA.

[2] Department of Chemical and Environmental Engineering, University of California, Riverside, CA 92521, USA.

## Abstract

Cellular uptake of nanoplastics is instrumental in their environmental accumulation and transfers to humans through the food chain. Despite extensive studies using spherical plastic nanoparticles, the influence of the morphological characteristics of environmentally released nanoplastics is understudied. Using dissipative particle dynamics (DPD) simulations, we modeled the interactions between hydrophobic nanotetrahedra and a cell membrane, featuring high shape anisotropy and large surface curvature seen for environmental nanoplastics. We observe robust uptake of nanotetrahedra with sharp vertices and edges by the lipid membrane. Two local energy minimum configurations of nanotetrahedra embedded in the membrane bilayer were identified for particles of large sizes. Further analysis of particle dynamics within the membrane shows that the two interaction states exhibit distinct translational and rotational dynamics in the directions normal and parallel to the plane of the membrane. The membrane confinement significantly arrests the out-of-plane motion, resulting in caged translation and subdiffusive rotation. While the in-plane diffusion remains Brownian, we find that the translational and rotational modes decouple from each other as nanotetrahedra size increases. The rotational diffusion decreases by a greater extent compared to the translational diffusion, deviating from the continuum theory predictions. These results provide fundamental insight into the shape effect on the nanoparticle dynamics in crowded lipid membranes.

*Email: xyong@binghamton.edu

**Introduction**

Synthetic polymers are ubiquitous in our society and daily life[1] and their immense utility simultaneously leads to a tremendous increase in plastic waste in landfills and the environment.[2–5] Nanoplastics can be defined as plastic fragments with sizes ranging from 1 nm – 1 μm and can exhibit colloidal behavior.[6,7] The majority of nanoplastics occurring in the environment are formed by fragmentation and degradation of microscopic and macroscopic plastic objects,[8–11] while a small fraction is directly released from industrial products and applications.[12,13] Much of the recent research regarding plastic pollution has been dedicated to microplastics. However, nanoplastics can behave very differently from their micron-sized counterparts. Small dimensions, large surface areas, and specific colloidal properties could facilitate the penetration of nanoplastics across the blood-brain and gut-blood barriers.[14] which presents an emerging threat to ecosystems and human health.[15,16]

The cell membrane is an important biological barrier for the uptake of nanoplastics. It is known that nanoparticles utilize two major pathways for transporting across the lipid membrane: direct penetration by passive diffusion and energy-consuming active translocation termed as endocytosis.[17–20] Both uptake pathways are governed by complex interactions of nanoparticles with the lipid bilayer, and the internalization process sensitively depends on particle size, shape, elasticity, chemical functionality, and surface charge.[21–24] Previous studies on nanoplastics predominantly used commercially available polystyrene latex nanospheres.[16] Such engineered models mimic nanoplastics sampled in the environment for size but do not capture highly irregular surface morphology resulting from natural weathering,[9,25,26] which can lead to biased results.[27,28] The lack of a comprehensive understanding of how shape anisotropy and large surface curvature influence cellular uptake of nanoplastics hampers our ability to assess the impact to the environment and humans accurately.

Experimentally, probing the transmembrane transport of nanoparticles challenges currently used sampling and detection techniques. Although microfluidic single-cell analysis[29,30] has provided much insight, the spatiotemporal resolutions of the reported experimental setups are still limited in *in situ* characterization of anisotropic nanoplastics interacting with the cell membrane in complex biological matrices. This leaves clear opportunities for computational simulations. For nanoparticles of sizes comparable to or smaller than the membranes, the finite thickness and lipid structure of the membrane become important in the interaction.[31] Particle-based simulation techniques have been widely used to model the penetration, accumulation, and spontaneous translocation of small nanoparticles in lipid membranes.[32–35] It has been shown that passive diffusion can result in the trapping of nanoplastics in the hydrophobic core of the bilayer.[36–38] A very recent coarse-grained molecular dynamics study using the MARTINI force field comprehensively investigated the biomembrane interaction of nanoplastics of five representative polymer types and three aging properties. The simulations revealed that interfacial processes, including nanoplastic translocation, shape transformation, and membrane perturbation, are governed by the competition of polymer–polymer and polymer–lipid interactions.[39] Despite extensive research on the interactions between anisotropic nanoparticles and cell membranes,[40–46] only a handful of previous studies explore the dynamics of colloid-membrane interaction in the context of environmentally relevant nanoplastics, possessing unique characteristics of irregular shape and specific surface chemistry. Moreover, most studies have focused on membrane insertion or translocation of a single particle or membrane-mediated assembly of multiple particles. To the extent of our knowledge, the detailed dynamics of anisotropic particles trapped within the membrane have not been explored on the quantitative level.

In this work, we explore the dynamics of tetrahedral nanoplastics interacting with a model plasma membrane using dissipative particle dynamics (DPD) simulations. We develop a new coarse-grained model of 1,2-dioleoyl-sn-glycero-3-phosphocholine (DOPC) bilayers (**Fig. 1a**) to accurately reproduce physical bilayer properties, which approximate those of eukaryotic plasma membranes.[31] The simulations elucidate the detailed dynamics of hydrophobic nanotetrahedra of varying sizes (**Fig. 1b**) interacting with the model membrane, focusing on the quantitative characterization of long-time dynamics. We find different interaction states of anisotropic nanotetrahedra in the membrane when the particle sizes become comparable to or larger than the membrane thickness. More interestingly, these states feature distinct translational and rotational dynamics of nanoparticles. Our results not only contribute to a comprehensive understanding of the shape effect in nanoplastic-membrane interaction but also provide insights into colloidal dynamics in crowded environments. Although this work focuses on nanotetrahedra with a general hydrophobic chemistry, which represent only a group of polymers within diverse nanoplastics detected in the environment,[47,48] the establishment of the new membrane model and dynamics analysis will serve as the starting point for studying irregular nanoplastics.

**Methods**

## A. Dissipative particle dynamics

DPD is a mesoscale method that has been widely used in simulating multiphase, multicomponent systems because of its computational efficiency and ease of implementation. The DPD model describes units of larger molecules or a group of smaller molecules as a single bead.[49,50] In contrast to Brownian dynamics, DPD includes explicit solvents and therefore is capable of capturing hydrodynamic interactions[51–53] important for membrane simulations. The DPD beads interact with each other through a pairwise conservative force supplemented by pairwise friction and random forces. The conservative force is a soft-core repulsion given by $\mathbf{F}_{ij}^C = a_{ij}(1 - r_{ij}/r_c)\mathbf{e}_{ij}$, where $a_{ij}$ is the maximum repulsion between beads $i$ and $j$; $r_{ij} = |\mathbf{r}_i - \mathbf{r}_j|$ is the inter-bead distance; and $\mathbf{e}_{ij} = (\mathbf{r}_i - \mathbf{r}_j)/|\mathbf{r}_i - \mathbf{r}_j|$ represents the force direction. $r_c$ is the cutoff radius of the interaction and is typically the same for all beads. The dissipative force $\mathbf{F}_{ij}^D$ is proportional to the relative velocity of two beads $\mathbf{v}_{ij} = \mathbf{v}_i - \mathbf{v}_j$, given by $\mathbf{F}_{ij}^D = -\eta w_D(r_{ij})(\mathbf{e}_{ij} \cdot \mathbf{v}_{ij})\mathbf{e}_{ij}$, where $\eta$ is a friction coefficient related to fluid viscosity. The random force is a Gaussian white noise $\mathbf{F}_{ij}^R = \sigma w_R(r_{ij})\xi_{ij}\mathbf{e}_{ij}$, where $\sigma$ is the noise amplitude and $\xi_{ij}$ is a random variable satisfying $\langle \xi_{ij}(t) \rangle = 0$ and $\langle \xi_{ij}(t)\xi_{i'j'}(t') \rangle = (\delta_{ii'}\delta_{jj'} + \delta_{ij'}\delta_{ji'})\delta(t - t')$. $w_D$ and $w_R$ are arbitrary weight functions dependent on the interparticle distance. They typically take the form of $[w_R(r_{ij})]^2 = w_D(r_{ij}) = (1 - r_{ij}/r_c)^2$ together with $\eta^2 = 2k_B T \sigma$ ($k_B$ is the Boltzmann constant) to satisfy the fluctuation-dissipation theorem for a system in equilibrium at temperature $T$. The bead motion is governed by the Newton's second law $m_i(d\mathbf{v}_i/dt) = \mathbf{f}_i$ with the total force $\mathbf{f}_i = \sum_{j \neq i}(\mathbf{F}_{ij}^C + \mathbf{F}_{ij}^D + \mathbf{F}_{ij}^R)$. The sum is carried out for all beads $j$ within the cutoff radius $r_c$ from bead $i$. As per convention of DPD simulations, the beads have the same mass $m_i = m$. The results are presented in reduced units using $r_c$ as the characteristic length scale and the energy of thermal fluctuation $k_B T$ at room temperature as the characteristic energy. The characteristic time scale is denoted as $\tau$. For simplicity, $r_c$, $m$, $k_B T$, and $\tau$ are all set to one. The equation of motion is integrated using the velocity-Verlet algorithm with a time step $\Delta t = 0.01\ \tau$. The value of $\eta$ is set at 4.5 to ensure numerical stability of the integrator and rapid equilibration of the system temperature.

## B. Model plasma membrane and tetrahedral nanoparticles

Inspired by the MARTINI model,[54] a DOPC molecule is composed of 14 beads with a $H_2E_2(C_5)_2$ architecture as shown in **Fig. 1a**. The zwitterionic phosphorylcholine group is represented by two hydrophilic beads (H). Two beads of intermediate hydrophilicity model the glycerol linking unit (E), with one of them connected to the head beads in a linear sequence. The oleoyl tails are composed of five hydrophobic beads (C) and attached to the two E beads, respectively. The central triplet (H-E-E) takes an off-linear configuration, which results in an h-shape lipid model. The lipid membrane is surrounded by an aqueous solvent composed of individual DPD beads. Each solvent (W) bead represents several water molecules. In this model, one interaction site corresponds to, on average, four heavy atoms.[54–56] This coarse-graining mapping not only preserves the asymmetric structure of phospholipids but also recognizes the chemical specificity of the glycerol ester moiety.

The bonded interactions are introduced between chemically connected sites, including bond and angle interactions. All adjacent lipid beads are attached together using harmonic springs with the potential $U_b(r_{ij}) = \frac{1}{2}K_b(r_{ij} - l_0)^2$ with a bond strength $K_b = 256 \ k_B T/r_c^2$ and an equilibrium bond length $l_0 = 0.5 \ r_c$. The bond strength is similar to the one used in the MARTINI model and the value equilibrium bond length is taken from previous DPD models.[57,58] The DPD interactions are excluded between bonded beads. Hydrocarbon chain stiffness is imposed by harmonic angle potentials applied on three adjacent C beads, given by $U_a(r_{ij}) = \frac{1}{2}K_a(\theta - \theta_0)^2$, where $K_a$ is the bending constant and $\theta_0$ is the equilibrium angle.[59] The DPD interactions between second nearest neighbors are not excluded. For single bonds, the angle potential parameters are $K_a = 20 \ k_B T$ and $\theta_0 = 180°$. The angles involving the cis double bond are assigned with higher bending stiffness $K_a = 36 \ k_B T$ with an equilibrium angle of 120°.[54] The triplets H-E-C and E-C-C use the same angle potentials as for the lipid tails, while the H-E-E triplet experiences a special angle potential for cis-bond bonds to maintain the h-shape topology.

Each plastic nanoparticle is modeled as a cluster of hydrophobic DPD beads (denoted by P), which are constrained to move as a rigid body. The constituent beads are assembled in a face-centered-cubic (fcc) structure. The corresponding number density of beads in a particle is set to $6 \ r_c^{-3}$ to prevent lipid and solvent beads from unphysically penetrating the particle.[58,60] The resulting lattice constant is $0.87 \ r_c$. Regular tetrahedrons of different sizes are carved from the fcc assembly with four faces being the (111), ($1\underline{1}\underline{1}$), ($\underline{1}1\underline{1}$), and ($\underline{1}\underline{1}1$) planes (see **Fig. 1b**). The edge lengths, $l_p$, of nanotetrahedra simulated in this work are 3.1, 5.6, 8.0, and 10.5 $r_c$. The total numbers of beads in each particle are 56, 220, 560, and 1140, respectively.

The dynamics of this system, including solvated lipid bilayers and nanoparticles, are largely governed by the hydrophilic and hydrophobic interactions between beads of different components. In DPD simulations, the chemical identity of each component is captured by the repulsion parameter of the conservative force $a_{ij}$. In principle, the self-interaction parameter can be related to the compressibility of the fluid, while the cross-interaction parameters can be obtained from the corresponding Flory-Huggins $\chi$ parameters and represent the hydrophobicity or hydrophilicity of the beads.[49,61] However, experimental $\chi$ values for a specific lipid and polymer in water are not all available.[62] Thus, the parameterization in this study is largely based on obtaining bilayers with structure and dynamic properties in good agreement with real DOPC bilayers. The set of conservative force parameters is tabulated in Table 1. For any two beads of the same type, the repulsion parameter is set to 25 $k_\text{B}T/r_c$. The interaction parameter between the hydrophilic (H and W) and hydrophobic (C and P) beads is set to 100 $k_\text{B}T/r_c$.[58,63] E beads have intermediate hydrophobicity with $a_{EC} = a_{EP} = 35\ k_\text{B}T/r_c$. This excess repulsion value corresponds approximately to $\chi_{EC} = 2$ introduced in Ref. [61]. This set of values leads to a bilayer with the desired structure and physical properties. We note that the compressibility of solvent will be lower than that of water at room temperature for $a_{WW} = 25\ k_\text{B}T/r_c$ with each bead representing four water molecules.[64] Nevertheless, we expect this discrepancy will not have a significant influence on the results of this work.

**C. Simulation setup**

For membrane-only simulations, a planar bilayer spanning the simulation box is constructed by placing DOPC molecules along the center *x-y* plane of the box (**Fig. 1a**). Periodic boundary conditions are applied to all three directions. Membranes of two sizes are simulated in this study. The characterization of membrane properties is performed using a small patch composed of 1152 lipid molecules. A large membrane with $N_{lp} = 4608$ lipids is used to probe the interaction with tetrahedral nanoplastics. This ensures the membrane is much larger than the particles and the membrane deformation upon nanoparticle association is not influenced by the periodic boundary conditions. Most DPD simulations, including ones for membrane simulations, are carried out in the *NVT* ensemble with fixed box dimensions. However, previous studies have shown that transverse finite-size effects influence the tension and thickness of the membrane simulated in the *NVT* ensemble.[62,65] To reduce the finite size effects, the membrane equilibration and accurate characterization of membrane properties are performed in the constant-normal pressure and constant-surface tension ($NP_N\gamma_s T$) ensemble.[66] The normal pressure is set to $P_N = 23.7\ k_\text{B}T/r_c^3$ to obtain the solvent density away from the bilayer $\rho = 3\ r_c^{-3}$, corresponding to standard DPD conditions.[49] We employ a simple Berendsen barostat with semi-isotropic coupling (i.e., the hydrostatic pressures in the $x$ and $y$ directions are coupled while the pressure in the z direction is separately controlled) to invoke an anisotropic pressure tensor in such a way that the desired values of normal pressure and surface tension are obtained.[66] The tensionless state of the model membrane can be readily modeled when $\gamma_s = 0\ k_\mathrm{B}T/r_c^2$ is imposed. The calculation of normal pressure and membrane tension is described in the results. For simplicity in results analysis, we perform particle interaction simulations in the $NVT$ ensemble after equilibrating the whole system in the $NP_N\gamma_sT$ ensemble.

System equilibration is performed for $1.5 \times 10^6$ time steps (15000 $\tau$). The production simulations of membrane structure analysis run for another $1.5 \times 10^6$ time steps, while the ones characterizing membrane and particle dynamics run $2 \times 10^6$ to $4 \times 10^6$ time steps. For the small membrane patch, the lateral dimensions of the box $L_x = L_y$ varies between 29.4 $r_c$ and 30.6 $r_c$ depending on the applied surface tension. The box height $L_z$ varies accordingly between 26.0 $r_c$ and 28.1 $r_c$ in the $NP_N\gamma_sT$ simulations. The box dimensions are fixed at $58.8 \times 58.8 \times 37.8\ r_c^3$ in the $NVT$ simulations of nanoparticle interaction with the large membrane patch. Independent simulations are performed to improve the statistics. Additional details of the simulation setup are given in the results and discussion. The simulations are performed using LAMMPS with standard packages.[67] Exemplary simulation scripts are available on GitHub (https://github.com/SaIL-Yong/nanotetrahedra).

**Results and Discussion**

**A. Improved DPD model for DOPC membrane mechanics**

Biological plasma membranes are multicomponent fluid bilayers containing a multitude of lipid species enriched with a significant fraction of cholesterol. The lipids and cholesterol are nonuniformly distributed in the membrane plane and between the two leaflets.[31] Instead of embracing the full complexity of realistic cell membranes,[68] we model the plasma membrane as a single-component DOPC bilayer in order to focus on membrane mechanics. Although DOPC is not the dominant lipid species in eukaryotic cells, the average properties of DOPC bilayers are closer to those of a plasma membrane than other common plasma membrane lipids (e.g., POPC).[31] Various coarse-grained models have been developed to simulate membrane lipids in DPD.[57,59,61,69] Here, we adopt features from different models to introduce a new $H_2E_2(C_5)_2$ molecular architecture (**Fig. 1a**) and parameterization, allowing the model membrane to reproduce key structural and mechanical properties of DOPC membranes.

We first conduct simulations of a membrane patch of 1152 lipids in the $NP_N\gamma_sT$ ensemble at the reference temperature $k_BT = 1$ and characterize the physical properties of a tensionless DOPC membrane modeled using the new coarse-grained parameterization. The membrane tension $\gamma_s$ can be calculated from the pressure tensor of the entire system given by $\gamma_s = \langle L_z[P_{zz} - 0.5(P_{xx} + P_{yy})]\rangle$ with the angular bracket representing the ensemble average. In addition, the membrane tension can also be obtained from the integral of lateral stress $\gamma_s = \int_{-\infty}^{+\infty} s(z)dz \approx \int_{-L_z/2}^{L_z/2} s(z)dz$ along the coordinate $z$ perpendicular to the membrane. $z = 0$ corresponds to the midplane of the membrane. The stress profile is given by $s(z) = P_N(z) - P_T(z)$, where $P_N(z)$ and $P_T(z)$ are the normal and tangential pressures across a plane parallel to the membrane through $z$. The normal pressure is the diagonal element of the local pressure tensor $P_N = P_{zz}$, which can be calculated by the Irving-Kirkwood formula.[70] Because the system is isotropic in the $x$ and $y$ directions, the tangential pressure is computed as the average of the corresponding diagonal elements $P_T = 0.5(P_{xx} + P_{yy})$. Away from the membrane surfaces, **Fig. 2a** shows that $P_N = P_T$ and both equal to the imposed bulk pressure of solvent at $\rho = 3\ r_c^{-3}$. For a system in mechanical equilibrium without external forces, the pressure tensor should satisfy the divergence-free condition $\nabla \cdot P = 0$. Specifically, this requirement leads to $\frac{\partial P_N}{\partial z} = 0$ and therefore constant normal pressure for a membrane system at equilibrium. The normal pressure profile remains largely flat throughout the system, which confirms that the equilibrium state of the membrane is obtained. The small variations observed in the membrane region are attributed to the noncentral force decomposition of multibody (i.e., angle) potentials in the stress calculation.[71] **Fig. 2b** plots the lateral stress profile across the membrane and its features can be understood with the help of the density profiles in **Fig. 2c**. The stress profile exhibits two peaks of positive stress at the membrane surfaces and negative stress in the hydrophobic core region. This stress distribution indicates that the hydrophilic head groups experience tensile stresses to balance the internal pressure of hydrophobic tails as a two-dimensional (2D) liquid under compression, resulting in the membrane's zero-tension state. In contrast to earlier DPD simulations that show a double-peak structure in the head group region,[57,72] the new coarse-grained model registers only a single peak of stress due to the incorporation of neutral E beads for the glycerol linkage. Compared to previous DPD models, the present stress profile is in better agreement with those obtained in coarse-grained MD simulations.[71]

The Berendsen barostat is efficient in equilibrating the system but is known not to sample the correct volume fluctuations. Considering that, we do not characterize bilayer mechanical properties based on natural membrane fluctuations. Instead, we employ a significantly more expensive method by simulating a series of stretched membranes, which are obtained via applying non-zero values of surface tension in the $NP_N\gamma_s T$ ensemble. The variations in the surface tension when the projected area per lipid $a_{pr} = 2L_xL_y/N_{lp}$ changes are shown in **Fig. 3**. We observe that the surface tension increases linearly as $a_{pr}$ increases when the membrane is close to the tensionless state. To further ensure the calculation is not affected by the Berendsen barostat, the deformed membranes are also simulated with constant areas in the absence of the barostat. The linear relationship between $\gamma_s$ and $a_{pr}$ is confirmed in another set of simulations in which the projected area per lipid is controlled by varying the lateral area $A = L_xL_y$ while keeping the number of lipid molecules and normal pressure fixed (denoted as $NP_NAT$ ensemble). We can determine the area expansion modulus $K_A$ as $\gamma_s \approx K_A(a_{pr} - a_0)/a_0$, where $a_0 \approx 1.505\ r_c^2$ is the area per lipid at zero surface tension. Based on the data in the linear regime from both the $NP_N\gamma_s T$ and $NP_NAT$ simulations, we obtain the value $K_A = 34.3\ k_B T/r_c^2$. The membrane bending rigidity can be further estimated using the theoretical relation[73] $\kappa = K_A h^2/48$. The membrane thickness $h$ is given by the distance between the two peaks of the density profile for the head groups in **Fig. 2c**. Substituting in $h = 5.4\ r_c$, the bending rigidity is approximately $20.8\ k_B T$ for our system.

The dynamic behavior of the membrane is explored by measuring the mean square displacement (MSD) of lipid molecules. Because the membrane spans the periodic simulation box in the $x$ and $y$ directions, we focus on the lipid transport in the lateral direction. The 2D MSD is calculated by the in-plane center of mass (COM) for each lipid $\langle \Delta r_{com}^2 \rangle \equiv \langle \Delta x_{com}^2 + \Delta y_{com}^2 \rangle$ with the angular bracket denoting the time and ensemble average, which is shown in **Fig. 4**. The lateral diffusion constant $D_T$ can be obtained accordingly through the Einstein relation $\langle \Delta r_{com}^2 \rangle = 4D_T t$ on long time scales. The linear fitting results in $D_T = 3.02 \times 10^{-3}\ r_c^2/\tau$. A similar value is obtained from the larger membrane with $N_{lp} = 4608$, confirming the calculation is not affected by the finite size effects. Notably, the solvent diffusivity is also measured in the same manner as a reference and the value is $D_w = 0.293\ r_c^2/\tau$, which is consistent with previous studies.[49,62] Compared with the lipid models used earlier simulations (e.g., $H_3(C_4)_2$ in Ref. 62), the lateral diffusivity of lipid in the new model is smaller. This difference is attributed to enhanced interactions between unsaturated tails with an increased length of 5 beads. Despite having a smaller diffusion constant, the model membrane is still in a fluid state.

The DPD length and time scales can be related to physical length and time scales by mapping the membrane properties to experimental results at room temperature. Experimental measurements of DOPC membranes[74] yield $a_0 = 0.724$ nm² at 30 °C and a thermal area expansivity of 0.0029 /deg, which give $a_0 = 0.714$ nm² at room temperature. From this experimental value, we obtain our simulation length scale to be $r_c = 0.689$ nm. The bilayer thickness in the simulations can then be calculated to be 3.72 nm, which agrees very well with the experimental value of 3.7 nm.[74] Taking the characteristic energy scale $k_BT = 4.11 \times 10^{-21}$ J, we find an area expansion modulus of 298 mN/m and a bending rigidity of $8.6 \times 10^{-20}$ J for the membrane in this work. Both elastic properties are also in good agreement with the experimental measurements,[75,76] testifying that the new model is self-consistent and accurately captures the physical properties of DOPC membrane. The DPD time scale $\tau$ can be estimated to be $\tau = 154$ ps from the lateral diffusion constant of DOPC molecules, which has been measured as 9.32 μm²/s.[77]

## B. Short-time interaction dynamics

When simulating nanoparticle membrane interactions, the common practice is to place the particle in the vicinity of the membrane surface and rely on the Brownian motion of the particle and thermal undulations of the membrane to initiate the particle-membrane contact. While mimicking experiments with cells incubated in nanoparticle suspensions, this approach can incur significant and unpredictable dwelling time in the present DPD simulations. To reduce the computational cost, the interaction between tetrahedral nanoplastics and the membrane is initiated by imposing a small impact velocity on the nanoparticle to drive it toward the membrane (**Fig. 5a**). To facilitate the analysis, the particle motion is constrained in the *x* and *y* directions in these simulations, so the particle is only allowed to move in the normal direction to the membrane (see **Video S1**). In contrast to the translation, there are no constraints applied on particle rotation because the orientation of anisotropic nanotetrahedra plays an important role in their interaction with the membrane.[78]

The membrane tension is initially zero before interacting with the particle. However, since the total projected membrane area is fixed in the *NVT* ensemble, the tension of the membrane with embedded nanotetrahedra is no longer strictly controlled at zero. We thus simulate the nanotetrahedra interactions with the large membrane patch having $N_{lp} = 4608$ lipids to reduce the influence of particle inclusion on the membrane tension. We also confirm that the deviated tension of the membrane with embedded particles of different sizes is within the range of fluctuations.

The particle velocity rapidly reduces as it moves to the membrane due to the viscous drag from the solvent. Because the surface-to-volume ratio increases as particle size decreases, smaller particles arrive at the membrane surface with less residual momenta given the same initial velocity (see **Fig. 5b**). The initial velocity $V_0$ is chosen to avoid inducing large perturbations to the membrane. We initially set an initial velocity $V_0 = 5\ r_c/\tau$ for all particles. However, the velocity of the smallest one of $l_p = 3.1\ r_c$ damps out so quickly that the particle cannot robustly reach the membrane. Therefore, we select a larger velocity $V_0 = 30\ r_c/\tau$ for the $l_p = 3.1\ r_c$ particle. Considering the upper surface of the membrane locates approximately at $z = 2.7\ r_c$, **Fig. 5a** shows that hydrophobic nanotetrahedra can readily overcome the energy barrier of the lipid head groups and be inserted into the hydrophobic tail region of the membrane. Note that the z-positions of the $l_p = 5.6\ r_c$ particles have larger deviations than particles of other sizes. This is because one independent simulation fails to establish the particle-membrane interaction, and the particle remains in the solvent (see **Fig. S1** for the detailed dynamics). Comparing nanotetrahedra of sizes 5.6, 8.0, and 10.5 $r_c$, the particle COMs are lower for bigger particles, indicating more extensive membrane deformation upon interaction as a result of larger particle momenta. Accordingly, bigger particles experience greater forces from the membrane at early times, as shown in **Fig. 5c**.

Independent simulations of the $l_p = 5.6\ r_c$ particles also provide evidence that hydrophobic nanotetrahedra can spontaneously insert into the membrane without the assistance of impact velocity. The z-positions of the particles in the second and third runs show plateau segments near $z_{COM} = 5\ r_c$ (**Fig. S1**). Concurrently, the particle velocity fluctuates around zero, indicating that the initial velocity has already damped out and the particle is undergoing Brownian diffusion. Following the plateau segment, the z position of the particle rapidly decreases toward zero, corresponding to the insertion process. In analyzing the detailed dynamics, we also reveal that the spontaneous insertion of nanotetrahedra is mediated by an interaction mode similar to the "corner attack" reported in previous work on hollow tetrahedra DNA nanostructures.[78] In particular, the nanotetrahedra cannot immediately insert when the particle approaches with one facet facing against the membrane surface. We observe that the particle rotates to position one of the vertices toward the hydrophilic head group layer, minimizing hydrophobic interactions and promoting penetration.

We further explore the interaction dynamics by monitoring the temporal evolution of the contact between nanotetrahedra and different regions of the membrane. The particle-membrane contact is quantified by the coordination number, defined as the number of neighbor beads of specific types within the cutoff distance $r_c$ from the particle. **Fig. 6a** shows that the coordinate numbers of $l_p = 3.1\ r_c$ particle with the lipid head and the solvent rapidly decrease to zero as the particle enter the membrane. Together with the increasing particle-tail contact, this indicates that the small hydrophobic particle is embedded well in the hydrophobic core of the membrane, minimizing total free energy by avoiding the interaction with hydrophilic components. The coordination number change also reveals the fast interaction and insertion of nanotetrahedra, occurring over the course of less than $100\ \tau$. As particle size increases, the hydrophobic core can no longer shield the entire particle. The particle vertices thus breach into the lipid head region and start interacting weakly with the surrounding solvent (**Fig. 6b-d**).

Interestingly, we observe an asymmetric configuration for particles of sizes 8.0 and 10.5 $r_c$, in which most of a nanotetrahedron is associated with the lower leaflet. The particles are oriented with a base plane parallel to the bilayer midplane and the 3-fold rotation axis through the opposite vertex perpendicular to the membrane (**Fig. S2c,d**). This inverted pyramid configuration results in significant exposure of the apex to the solvent, which entails a large enthalpic penalty. We contend that this state is metastable and kinetically trapped due to the impact of nanoparticles from the top of the membrane (see **Video S2**). Indeed, **Fig. S2c** shows that the $l_p = 8.0\ r_c$ particle is symmetrically embedded in the membrane in one independent run, with its COM approximately at the midplane of the bilayer. In contrast to the asymmetric insertion, this configuration is conceptually analogous to the interaction state observed for small particles. **Fig. 7** presents a direct comparison between the contact dynamics of nanotetrahedra of size 8.0 $r_c$ with the membrane in the two interaction modes. The particle in the symmetric mode respectively increases and reduces its contact with the membrane core and the solvent. Hence, this configuration should possess a lower enthalpy than the asymmetric configuration.

Knowing different embedding configurations, we continue running the simulations for an additional $2 \times 10^6$ time steps (i.e., 20,000 $\tau$) to probe the occurrence of two configurations.

The transition from asymmetric to symmetric configurations has been observed, confirming that the symmetric one is energetically favorable. However, both symmetric and asymmetric configurations can be obtained for particles of sizes 8.0 and 10.5 $r_c$ as the final states in the long runs of 20,000 $\tau$. This behavior suggests a considerable energy barrier between the two states. We therefore explore them as two different interaction states in the following study of long-time dynamics. Notably, we observe in multiple independent runs that the embedded nanotetrahedra adopts the asymmetric configuration more frequently.

**C. Nanoplastic translation and rotation in the membrane**

We quantitatively characterize the long-time dynamics of plastic nanoparticles within the membrane, which could influence their spatiotemporal distribution and effective particle-particle interactions. Although the translational dynamics can be easily probed through tracking the particle COM, additional orientation vectors need to be introduced to explore the rotational degrees of freedom of anisotropic particles and measure the orientation changes. Considering the tetrahedral symmetry, we introduce four normalized unit vectors $\hat{\mathbf{p}}_i(t)$ pointing from the centroid to four vertices, as shown in **Fig. 8**. The instantaneous particle orientation can thus be inferred by the polar ($\theta$) and azimuthal ($\phi$) angles of $\hat{\mathbf{p}}_i(t)$. **Fig. 9** plots the angular distributions of the orientation vectors for nanotetrahedra with increasing sizes to elucidate the size effect on the rotational dynamics. Here, $\theta = 0$ and $\phi = 0$ align with the $+z$ and $+x$ directions of the simulation box, respectively. The azimuthal angle distributions of small particles of sizes 3.1 and 5.6 $r_c$ are approximately uniform. Their polar angle distributions are largely consistent with uniformly random orientations (**Fig. S3**). **Figs. S4 and S5** further show that the state points corresponding to a specific vector spread throughout the entire $\theta$-$\phi$ plane. These results imply that small particles in the membrane have nearly random orientations and surrounding lipids do not hinder their rotation.

For $l_p = 8.0$ and 10.5 $r_c$ particles, the dynamics of the two embedding states are remarkably different. The symmetric state populates polar angles of 55° and 125°, corresponding to half tetrahedral angles from the $+z$ and $-z$ directions. This polar angle distribution confirms that the hydrophobic nanoparticle locates at the bilayer midplane to minimize exposure to the solvent and maximize the interactions with lipid tails. In contrast, the polar angle distributions for the asymmetric state show high density around 70° and 180°, consistent with the inverted pyramid orientation. The azimuthal angle distributions capturing the in-plane rotation (with the axis normal to the membrane) also differ between the two states.

The symmetric state exhibits pronounced nonuniform distributions, while the particle orientations in the asymmetric state are relatively isotropic. **Figs. S6 and S7** further demonstrate that the in-plane rotation of these nanotetrahedra in the symmetric configuration is arrested.

The nanoparticle rotation will cause a rotation of $\hat{\mathbf{p}}_i(t)$. However, an angular displacement as $\hat{\mathbf{p}}_i(t) - \hat{\mathbf{p}}_i(0)$ as a simple analog to the translation displacement $\mathbf{r}(t) - \mathbf{r}(0)$ will be bounded by the surface of a unit sphere, as shown in **Video S3,** and fail to capture full particle rotations.[79] To properly analyze the rotational dynamics in 3D, an angular displacement vector $\Delta\boldsymbol{\varphi}_i(t)$ can be defined for $i$th orientation vector $\hat{\mathbf{p}}_i$ in a time interval $[t, t+dt]$. Here, the magnitude of $\Delta\boldsymbol{\varphi}_i(t)$ quantifies the angle subtended by $\hat{\mathbf{p}}_i$ during the interval, such that $|\Delta\boldsymbol{\varphi}_i(t)| \equiv \cos^{-1}[\hat{\mathbf{p}}_i(t) \cdot \hat{\mathbf{p}}_i(t+dt)]$. The direction of $\Delta\boldsymbol{\varphi}_i(t)$ aligns with the rotation axis, given by $\hat{\mathbf{p}}_i(t) \times \hat{\mathbf{p}}_i(t+dt)$. Given $\Delta\boldsymbol{\varphi}_i(t)$, a total vector rotational displacement can be defined through the integral $\boldsymbol{\varphi}_i(t) = \int_o^t \Delta\boldsymbol{\varphi}_i(t')dt'$.[80,81] This vector allows one to define a trajectory in a rotation space $(\varphi_x, \varphi_y, \varphi_z)$, representing cumulative rotations about the Cartesian axes. In analogy to the translational MSD, an unbounded rotational mean square displacement (RMSD) for each orientation vector is given by $\langle \varphi_i^2(\Delta t)\rangle \equiv \langle |\boldsymbol{\varphi}_i(t+\Delta t) - \boldsymbol{\varphi}_i(t)|^2\rangle$. An average RMSD of nanotetrahedron can be obtained by future averaging among four orientation vectors $\langle \varphi^2(\Delta t)\rangle = \frac{1}{4}\sum_{i=1}^{4}\langle \varphi_i^2(\Delta t)\rangle$ in different trajectories from independent runs.

Considering the geometric constraints of the membrane, we separately probe the lateral mode (parallel to the plane of the membrane) and the transverse mode (perpendicular to the membrane) of translation and rotation. MSD and RMSD of nanoplastics in the membrane show two regimes (**Fig. 10**). In short times, the translational and rotational dynamics exhibit a power-law scaling to simulation time with exponents of 2 in all directions. This behavior indicates that the short-time translation and rotation of nanoparticles are ballistic and not influenced by particle size. The MSD in the $z$ direction (Fig. 9b) plateaus for long times, indicating that the particle is caged in the normal direction by the membrane. Parallel to the membrane, the lateral mode of particle translation exhibits Brownian diffusion with a linear scaling relation on long-time scales. Despite being in a crowded environment, the translation motion of nanotetrahedra shows no fundamental differences from that in the solvent, evidenced by the transition between the ballistic and diffusive regimes.

**Fig. 10** also reveals that small nanotetrahedra undergo Brownian rotational diffusion regardless of whether the rotational axis aligns parallel or normal to the membrane. As particle size increases, the transverse rotation mode is gradually suppressed, showing a pronounced subdiffusive behavior. Notably, the rotation in the membrane plane remains diffusive even for the largest particle of size 10.5 in both embedding states. We can obtain in-plane translational and rotational diffusion coefficients for the measurements of the MSD and RMSD. Considering their corresponding degrees of freedom, the lateral translational and rotational diffusivities are given by $\lim_{\Delta t \to \infty} \langle x^2(\Delta t) + y^2(\Delta t) \rangle = 4D_t^L \Delta t$ and $\lim_{\Delta t \to \infty} \langle \varphi_z^2(\Delta t) \rangle = 2D_r^L \Delta t$, respectively. **Fig. 11** shows that both diffusion coefficients decrease as particle size increases when nanotetrahedra are well immersed in the bilayer core region. However, the asymmetric configuration observed for large particles has much larger diffusivities for both in-plane translation and rotation than the symmetric configuration. The hindered dynamics in the symmetric state are attributed to the much higher viscosity of the surrounding membrane lipids than the solvent. Interestingly, the symmetric state exhibits higher mobility than the asymmetric state for out-of-plane rotating, even though the particle strongly interacts with lipids (see the inset of **Fig. 9d**). When asymmetrically inserted, the nanotetrahedron significantly deforms the membrane-solvent interface (**Fig. S2d**) and thus experiences a capillary force that restrains its out-of-plane rotation.[82,83]

**D. Discussion**

Compared with extensively studied nanospheres, hydrophobic nanotetrahedra with a high degree of shape anisotropy and significant surface curvature variation exhibit distinct behaviors when interacting with the biomembranes as model nanoplastics. In contrast to spherical nanoparticles encapsulated symmetrically in the membrane,[84,85] nanotetrahedra present a unique asymmetric configuration biased toward the inner leaflet (with respect to the insertion direction) when the particle size is comparable or larger than the membrane thickness. This uptake state is achieved by the partial translocation ensuing from the particle impact, as demonstrated in the dynamic simulations. This configuration is reminiscent of the one documented for nanotetrahedra interacting with a pulmonary surfactant layer.[43] Although the particles are subject to unfavorable enthalpic interactions in this configuration, their long-time dynamics surprisingly reveal high in-plane translational and rotational diffusivities. The entropy gain could offset the enthalpic penalty, rendering a small difference in the free energies of the asymmetric and symmetric configurations. Nevertheless, because the simulations readily capture the asymmetric-to-symmetric transition, we believe the symmetric configuration still corresponds to the global energy minimal state.

Although the diffusion of proteins and lipids in crowded lipid membranes has been studied,[86–89] our understanding of nanoparticle diffusion, particularly anisotropic ones, in these confined media is still limited. The diffusion behavior of spherical nanoparticles in the membrane is Brownian and can be well described by the hydrodynamic Saffman-Delbrück model when the particle size is small.[90] In contrast to the diffusion in the bulk solution described by the Stokes-Einstein and Stokes-Einstein-Debye relations, the translational diffusion of membrane-trapped particles depends weakly on the particle radius through a logarithmic relation $D_t^L \propto \ln(R^{-1})$, while the rotational diffusivity shows a strong algebraic size dependence $D_r^L \propto R^{-2}$ .[91] However, numerical simulations have shown significant deviations of $D_t^L$ from the logarithm decrease when the particle size exceeds a critical value.[92] The theoretical models also fail when particle surface charge are considered explicitly.[93]

We find that the translational and rotational dynamics of nanotetradehra depend apparently on the embedding configuration, but the relationship between them has not been uncovered. **Fig. 11** shows that as the particle is increasingly caged by the membrane when the size increases, both translational and rotation diffusion become slow. Unfortunately, due to limited particle sizes, we are not able to characterize the size-dependence diffusion of nanotetrahedra. However, the ratio of $D_t^L/D_r^L$ shows a size dependence stronger than the quadratic scaling $R^2$ (see **Fig. 11** inset). The diffusion of highly anisotropic particles in the lipid membrane is significantly different from that of spherical particles and cannot be explained by the Saffman-Delbrück theory. The translational and rotational modes of diffusion are clearly decoupled for the nanotetrahedra. The anomalous decoupling suggests that the crowded membrane environment renders the rotational dynamics of anisotropic particles more frustrated than the translational dynamics. We speculate that this behavior may be attributed to different effects of surface tension on translational and rotational diffusion, which will be investigated in future studies. It is also noteworthy that the effects of particle-membrane association on the dynamics are asymmetric. While the lipids show a pronounced influence on particle dynamics, the presence of nanotetrahedra does not affect lipid diffusion (**Fig. S8**) despite can induce considerable deformation to the membrane.

**Conclusions**

To provide physical insight into the cellular uptake of nanoplastics, we performed coarse-grained computer simulations to probe the effects of surface morphology on the interactions between hydrophobic tetrahedral nanoparticles and a lipid membrane on the mesoscale. The study improved the DPD model of the DOPC bilayer to accurately capture the membrane's physical properties, which serves as a model for eukaryotic plasma membranes. Using the new membrane model, the dynamics of nanotetrahedra interacting with the membrane were elucidated on different time scales. Taking advantage of its sharp vertices, hydrophobic nanotetrahedra can penetrate the hydrophilic surface layer and spontaneously insert into the membrane core through the "corner attack" mechanism. As its size increases, the anisotropic tetrahedra adopt an asymmetric configuration in the membrane in addition to being symmetrically encapsulated by the bilayer. In this configuration, nanotetrahedra are located in the lower leaflet with one of its vertices significantly exposed to the solvent. The characterization of particle translation and rotation reveals clear size-dependent dynamics. While the out-of-plane motion is confined by the membrane, the in-plane dynamics show normal diffusion for both translation and rotation in the two configurations. The asymmetric configuration exhibits higher in-plane diffusivity due to reduced interactions with lipids. Although diffusion is Brownian, the in-plane translation and rotation show an anomalous decoupling in the crowded lipid environment as particle size increases, in which the rotation is more strongly suppressed. Our findings shed light onto the unusual dynamics of anisotropic nanoparticles interacting with biomembranes and have important implications in the self-assembly and collective dynamics of nanoplastics during the cellular uptake.

**Acknowledgments**

X. Y. and K. D. acknowledge funding from the National Science Foundation for supporting this work through awards 2034855 and 2035623. Computing time was provided by the Center for Functional Nanomaterials, which is a U.S. DOE Office of Science Facility, at Brookhaven National Laboratory under Contract No. DESC0012704.

**Supporting Information:** Additional figures showing detailed interaction dynamics for each independent runs. Comparison between orientations of $l_p = 3.1\, r_c$ and $l_p = 5.6\, r_c$ nanotetrahedra in the membrane with ideal uniformly random orientations. Detailed distributions of the orientation vectors of different sized nanotetradra. Lipid dynamics in the membranes with embedded nanotetrahdra of different sizes. Supplementary videos of nanotetradra impact and rotational trajectories in the membrane.

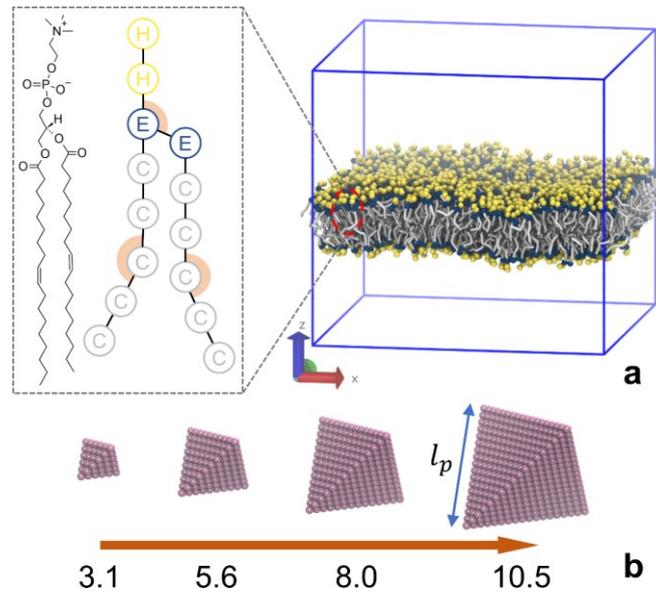

Fig. 1. Coarse-grained computational model of a model plasma membrane and tetrahedral nanoplastics. (a) (Left) molecular structure and coarse-grained mapping of 1,2-dioleoyl-sn-glycero-3-phosphocholine (DOPC). The angles involving the glycerol backbone and cis-unsaturated bonds are marked in orange. (Right) DPD simulation snapshot of an equilibrium DOPC bilayer membrane ($N_{lp} = 1152$) at the zero-tension state. The zwitterionic head groups, glycerol ester moiety, and hydrocarbon tails are represented by yellow, blue, and grey beads. Surrounding solvent beads are not shown for clarity. (b) Nanotetradrons of various sizes composed of DPD beads arranged in an fcc lattice.

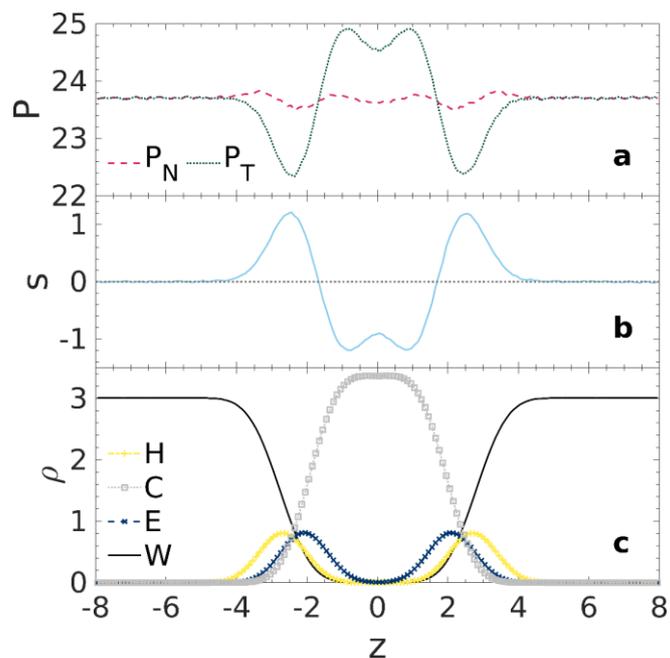

Fig. 2. Local physical properties of the membrane system. (a) Normal and tangential pressures as a function of $z$ for the tensionless DOPC membrane with $N_{lp} = 1152$. (b) The corresponding stress profile across the membrane. The dotted line denotes zero stress. Positive values of stress indicate tension and negative stresses indicate compression. (c). Transmembrane number density profiles of lipid head and tail groups, glycerol backbone, and water. The color scheme is the same as in Fig. 1a. The calculation is performed using trajectories per 100 time steps during the last $1 \times 10^6$ time steps of the production run ($1 \times 10^4$ frames in total).

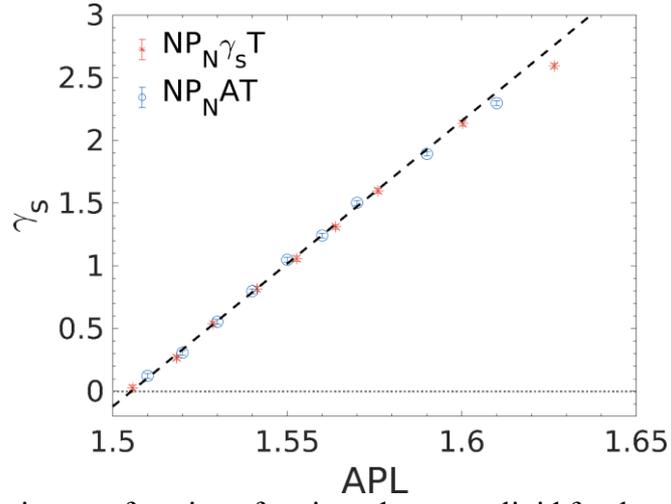

Fig. 3. Membrane tension as a function of projected area per lipid for the simulations performed in two different ensembles. Error bars represent the standard deviation. The dashed line represents the least-square linear fitting to the simulation data for $\gamma_s < 1.2\ k_\mathrm{B}T/r_c^2$. The dotted line marks $\gamma_s = 0$. The intersection of two lines denotes the zero-tension area per lipid $a_0$.

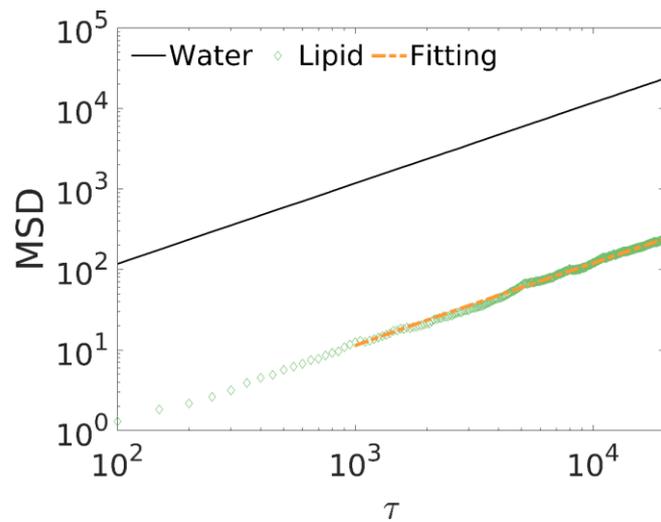

Fig. 4. Mean square displacement of water and lipids in the direction parallel to the membrane. The slope of the linear fitting gives the lateral diffusion constant of lipids.

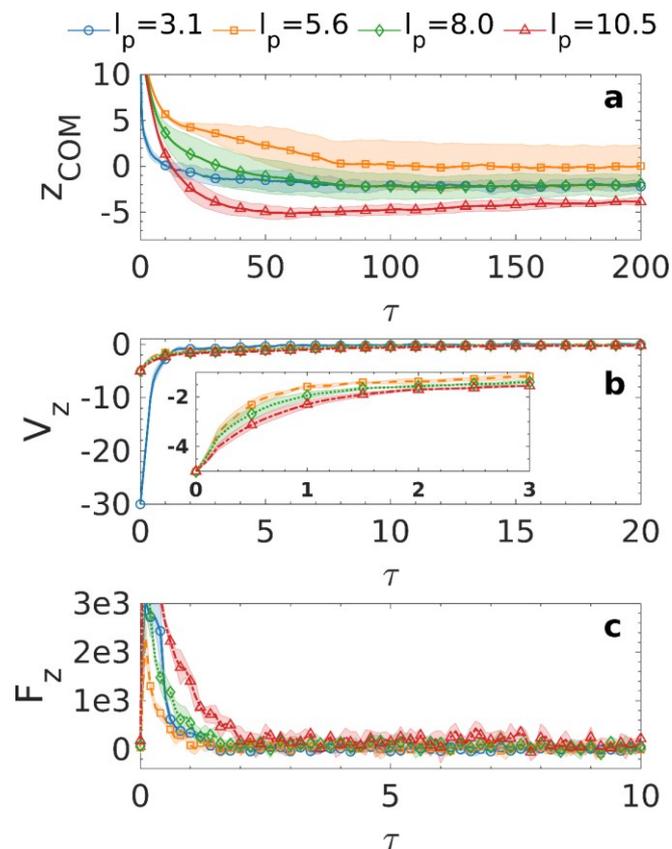

Fig. 5. Early-stage interaction dynamics of hydrophobic nanotetrahedra the DOPC membrane. (a) the *z*-position of the particle's center of mass, (b) the particle velocity in the *z* direction, and (c) the *z*-component of the force acting on the particle of different sizes (3.1, 5.6, 8.0, and 10.5 $r_c$) interacting with as functions of simulation time. The shaded error region represents the standard deviations of five independent runs. The inset in (b) shows the zoom-in view of the velocity variations for particles having the same initial velocity.

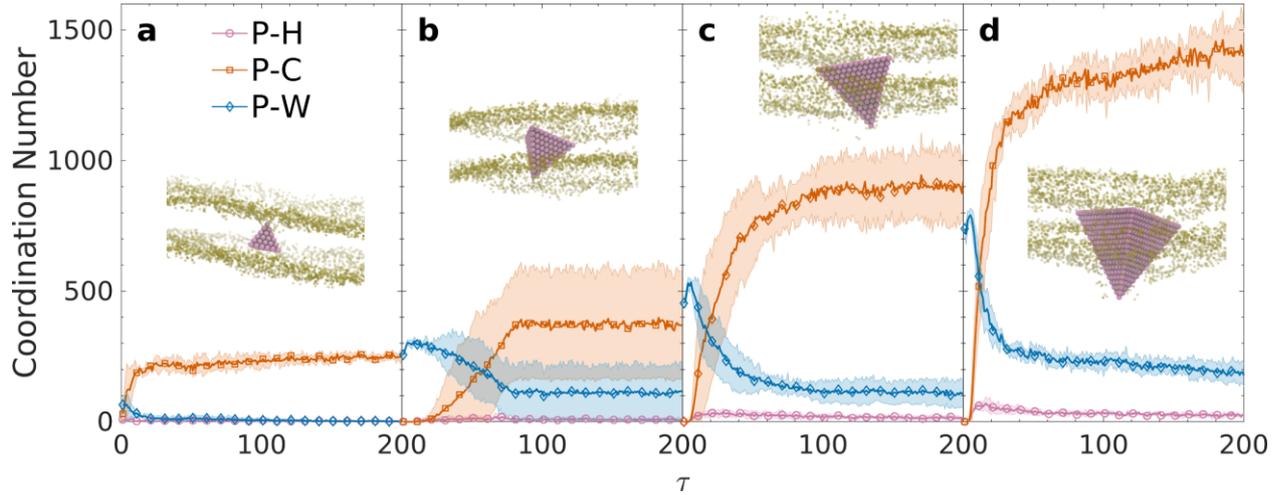

Fig. 6. Early-state contact dynamics of nanotetrahedra with the membrane. Time evolution of the coordination numbers of particle beads (P) interacting with lipid head (H), lipid tail (C), and the solvent (W) beads for nanotetrahedra of sizes (a) 3.1, (b) 5.6, (c) 8.0, and (d) 10.5 $r_c$ during the early stages of interaction. The shaded error region represents the standard deviations of five independent runs. Similar to Fig. 5, the large deviations for $l_p = 5.6\, r_c$ particle are attributed to unsuccessful interaction in one independent runs (see Fig. S2 for details). The insets show the representative positions and orientations of nanotetrahedra in the membrane. Here, the membrane surfaces are illustrated by yellow lipid head beads and the nanoparticles are represented by pink beads. The hydrophobic core of the membrane and the solvent are not displayed for clarity.

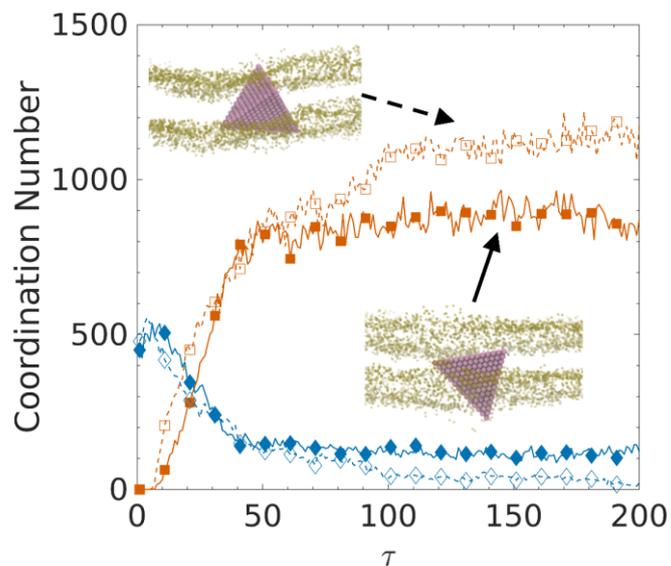

Fig. 7. Comparison between the contact dynamics of nanotetrahedra of size 8.0 $r_c$ interacting with the membrane in the two different modes. The color scheme is the same as Fig. 6, with orange and blue corresponding to the particle-tail (P-C) and particle-solvent (P-W) interactions, respectively. The solid lines and closed symbols show the contact dynamics in the asymmetric embedding configuration, while the dashed lines and open symbols represent the symmetric embedding configuration. The insets are the simulation snapshots of the two interaction modes.

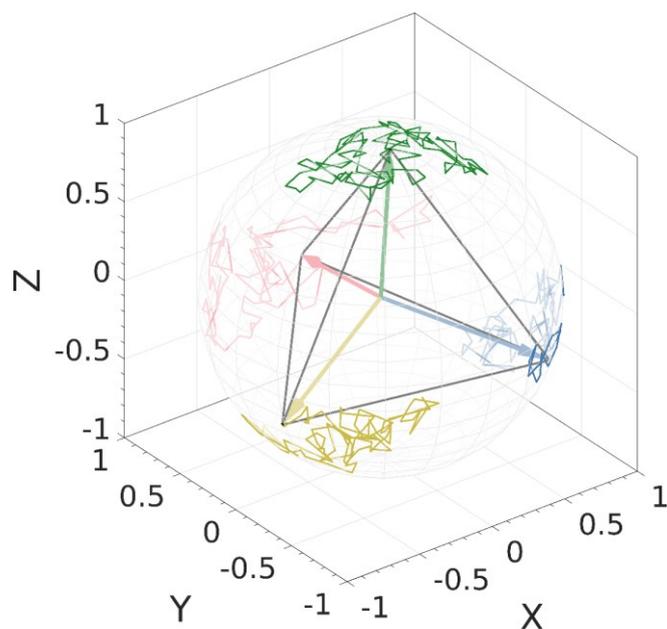

Fig. 8. Trajectories of four orientation vectors (colored differently) of a rotating nanotetrahedron projected onto the surface of a unit sphere. The corresponding simulation time is $100\ \tau$.

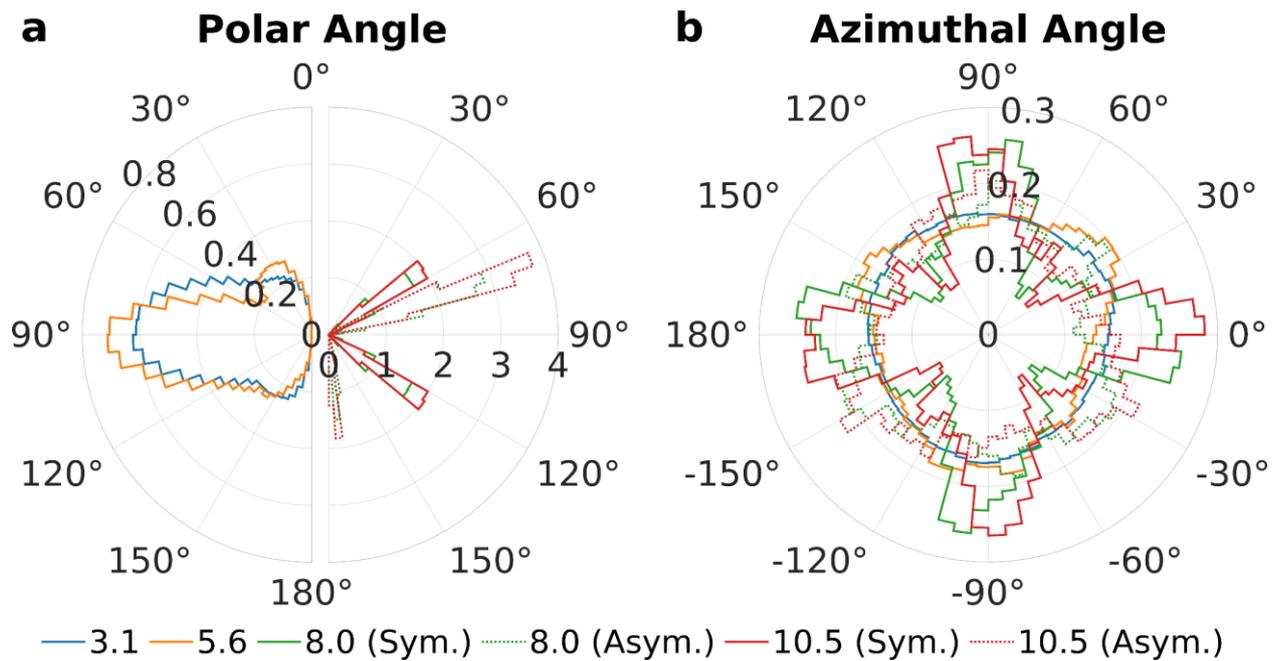

Fig. 9. Nanotetrahedra orientations in the membrane. Polar histograms of the probability densities of (a) polar angles and (b) azimuthal angles of the orientation vectors for particles of of different sizes (3.1, 5.6, 8.0, and 10.5 $r_c$). The symmetric and asymmetric embedding configurations are distinguished for large particles. The data is sampled from four independent runs for each particle size and state.

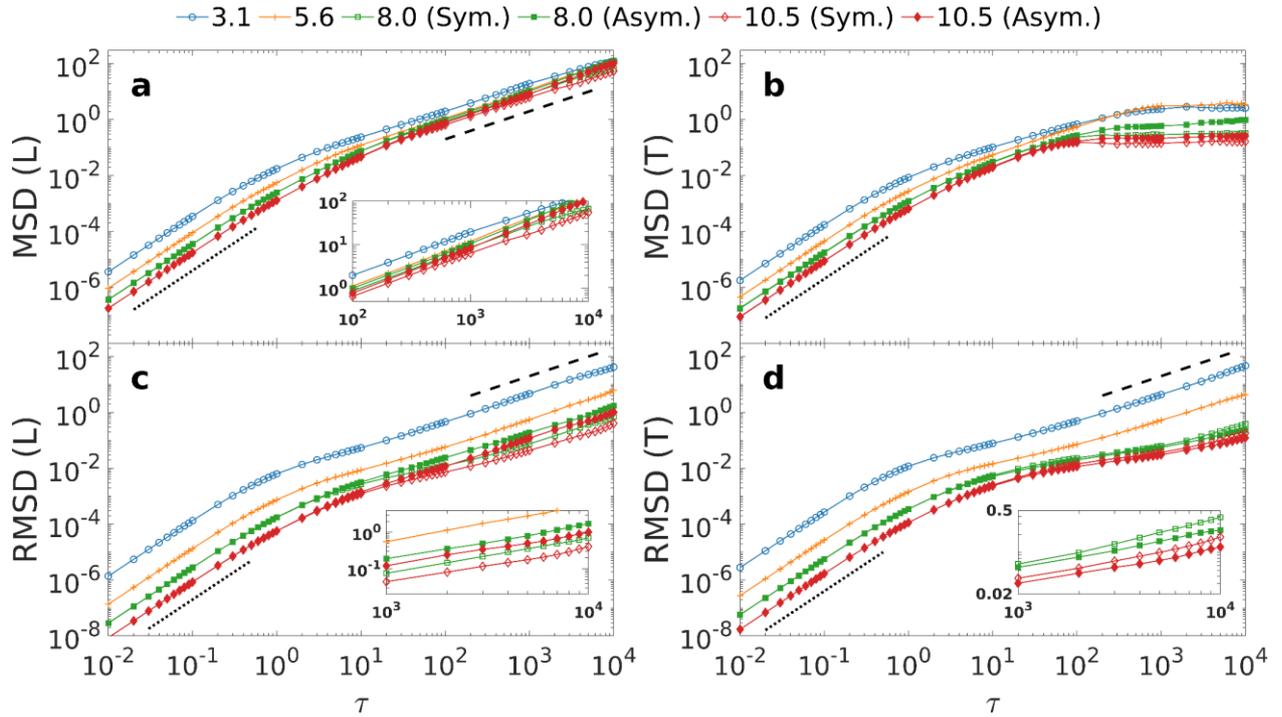

Fig. 10. Translation and rotational mobility for membrane-associated nanotetrahedra. (a) MSD in the $x$ and $y$ directions, parallel to the membrane. (b) MSD in the $z$ direction, along the membrane normal. (c) RMSD about the $z$-axis. (d) RMSD about the $x$ and $y$ axes. The solid lines are guides for the eye. The dotted lines have slope 2 and the dashed lines have slope 1. The insets are the zoomed-in views of the long-time behavior. The results are the average of four independent runs.

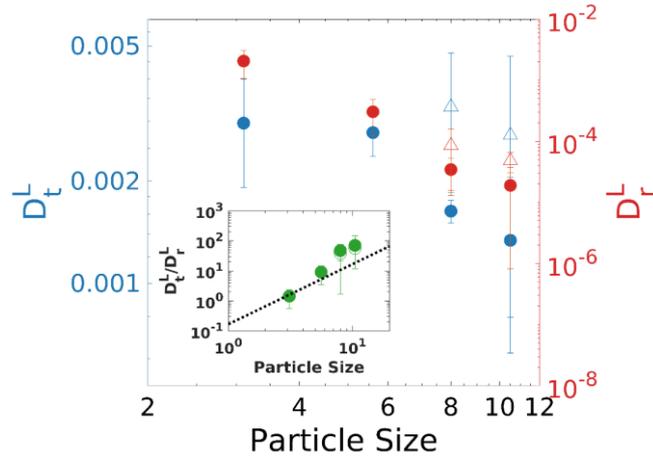

Fig. 11. Diffusion coefficients of nanotetrahedra of different sizes corresponding to translation and rotation in the plane of membrane. Blue and red circles are the translational (left axis) and rotational (right axis) diffusion coefficients for the symmetric embedding configuration. Open symbols mark the data for the asymmetric embedding configuration. The inset plots the ratio between diffusion coefficients. The dotted line has slope 2, indicating a scaling relation of $D_t^L/D_r^L \sim l_p^2$. The error bars show the standard deviations of four independent runs.

| $a_{ij}$ | H | E | C | W | P |
|---|---|---|---|---|---|
| H | 25 | 25 | 100 | 25 | 100 |
| E | - | 25 | 35 | 25 | 35 |
| C | - | - | 25 | 100 | 25 |
| W | - | - | - | 25 | 100 |
| P | | | | | 25 |

Table 1. Repulsion parameters of DPD conservative force for DOPC membranes and nanotetrahedra (in units of $k_\mathrm{B}T/r_c$).

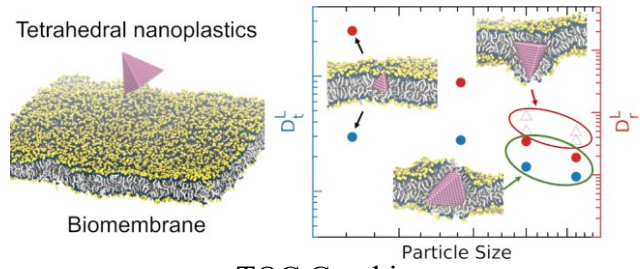

TOC Graphic